\documentclass[a4paper]{article}
\usepackage[english]{babel}
\usepackage[utf8x]{inputenc}
\usepackage[T1]{fontenc}

\usepackage[letterpaper,top=1.5cm,bottom=2cm,left=1.5cm,right=1.5cm,marginparwidth=1.75cm]{geometry}

\usepackage{amsmath}
\usepackage{graphicx}
\usepackage[colorinlistoftodos]{todonotes}
\usepackage[colorlinks=true, allcolors=blue]{hyperref}
\usepackage{siunitx}
\usepackage{authblk}
\usepackage{xcolor}
\usepackage{nameref}
\usepackage{afterpage}

\usepackage{caption}
\usepackage{algorithm}

\usepackage[]{natbib}
\newcommand{\btheta}{\boldsymbol{\theta} }

\usepackage{setspace}
\title{Bayesian Uncertainty Quantification for Systems Biology Models Parameterized Using Qualitative Data}
\author{Eshan D. Mitra}
\author{William S. Hlavacek \thanks{Corresponding author. wish@lanl.gov}}
\affil{Theoretical Biology and Biophysics Group, Theoretical Division, Los Alamos National Laboratory, Los Alamos, NM, United States}

\begin{document}

\maketitle

\clearpage

\section*{Abstract}

\textbf{Motivation:} Recent work has demonstrated the feasibility of using non-numerical, qualitative data to parameterize mathematical models. However, uncertainty quantification (UQ) of such parameterized models has remained challenging because of a lack of a statistical interpretation of the objective functions used in optimization.\\
\textbf{Results:} We formulated likelihood functions suitable for performing Bayesian UQ using qualitative data or a combination of qualitative and quantitative data. To demonstrate the resulting UQ capabilities, we analyzed a published model for IgE receptor signaling using synthetic qualitative and quantitative datasets. Remarkably, estimates of parameter values derived from the qualitative data were nearly as consistent with the assumed ground-truth parameter values as estimates derived from the lower throughput quantitative data. These results provide further motivation for leveraging qualitative data in biological modeling.\\
\textbf{Availability:} The likelihood functions presented here are implemented in a new release of PyBioNetFit, an open-source application for analyzing SBML- and BNGL-formatted models, available online at \url{www.github.com/lanl/PyBNF}.\\


\clearpage

\section{Introduction}

Mathematical models of the dynamics of cellular networks, such as those defined using BioNetGen Language (BNGL) \citep{Faeder2009} or Systems Biology Markup Language (SBML) \citep{Hucka2003}, require parameterization for consistency with experimental data. Conventional approaches use quantitative data such as time courses and dose-response curves to parameterize models. We and others have demonstrated that it is also possible to use non-numerical, qualitative data in automated model parameterization \citep{Oguz2013, Pargett2013, Pargett2014, Mitra2018a}. Our demonstration \citep{Mitra2018a} used qualitative data in combination with quantitative data. 

In the method of \cite{Mitra2018a}, the available qualitative data are used to formulate inequality constraints on outputs of a model. Parameterization is performed by minimizing a sum of static penalty functions \citep{Smith1997} derived from the inequalities. Given a list of $n$ inequalities of the form $g_i<0$ for $i=1,...,n$, where the $g_i$ are functions of model outputs, the objective function is defined as
\begin{equation}
    \sum_{i=1}^n C_i\cdot\max(0, g_i)
    \label{eq:static}
\end{equation}

Static penalty functions have long been used in the field of constrained optimization \citep{Smith1997}. Each violated inequality contributes to the objective function a quantity equal to a distance from constraint satisfaction (e.g., the absolute difference between the left-hand side and right-hand side of the inequality), multiplied by a problem-specific constant weight $C_i$. The objective function of Equation \ref{eq:static} becomes smaller as inequalities move closer to satisfaction, thus guiding an optimization algorithm toward a solution satisfying more of the inequalities. In the study of \cite{Mitra2018a}, the approach proved effective in obtaining a reasonable point estimate for the parameters of a 153-parameter model of yeast cell cycle control developed by Tyson and coworkers \citep{Chen2000,Chen2004,Csikasz-Nagy2006,Oguz2013,Kraikivski2015}, which had previously been parameterized by hand tuning.  

The static penalty function approach has limitations. Most notably, the approach requires choosing problem-specific weights $C_i$ for the objective function. Although heuristics exist to make reasonable choices for the weights \citep{Mitra2018a}, there is no rigorous method to do so. A related challenge in using qualitative data is performing uncertainty quantification (UQ). 

Bayesian UQ (described in many studies, such as \cite{Kozer2013} and \cite{Klinke2009}) is a valuable approach that generates the multivariate posterior probability distribution of model parameters given data. This distribution can be used for several types of analyses. 1) The marginal distribution of each parameter can be examined to find the most likely value of that parameter and a credible interval. 2) Marginal distributions of pairs of parameters can be examined to determine which parameters are correlated. 3) Prediction uncertainty can be quantified by running simulations using parameter sets drawn from the distribution. Unfortunately, meaningful Bayesian UQ cannot be performed for models parameterized using qualitative data and penalty function-based optimization, because the penalty functions are heuristics. They are not grounded in statistical modeling. 

Here, we present likelihood functions that can be used in parameterization and UQ problems incorporating both qualitative and quantitative data. We first present a likelihood function that can be used with binary categorical data, and then a more general form to use with ordinal data comprising three or more categories. We implemented the option to use these likelihood functions in fitting and in Bayesian UQ in our software PyBioNetFit \citep{Mitra2019a}. We built on existing PyBioNetFit support for qualitative data, which previously allowed only the static penalty function approach. In the first section of Results, we derive the new likelihood functions, which have similarities to both the chi squared likelihood function commonly used in curve fitting with quantitative data, and the logistic function commonly used to model classification error in machine learning. In the second section, we describe how we have added support for the new likelihood functions in PyBioNetFit and provide a guide to using them in optimization and UQ. In the third section, we provide an example application of the new software features. This example shows that qualitative datasets are potentially valuable resources for biological modeling.

\section{Methods}

Likelihood functions presented in Results were implemented as options in PyBioNetFit v1.1.0, available online at \url{https://github.com/lanl/pybnf}. PyBioNetFit supersedes the earlier BioNetFit \citep{Thomas2016, Hlavacek2018}.

To illustrate use of the new functionality, we configured and solved an example UQ problem (described in Section \ref{sec:application}) using PyBioNetFit v1.1.0. Configuration, model, and synthetic data files used for this example are available online (\url{https://github.com/RuleWorld/RuleHub/tree/2019Aug27/Contributed/Mitra2019Likelihood}). The model that we used has been published in BNGL format \citep{Faeder2009} in earlier work \citep{Harmon2017}. We took the published parameterization to be the ground truth. We adapted the simulation commands included in the BNGL file to produce degranulation outputs for specific conditions, as appropriate for our synthetic datasets described below. 

We considered 11 instances of the problem using different qualitative and quantitative datasets. To generate synthetic quantitative data, we simulated the model with the assumed ground-truth parameterization, and added Gaussian noise to the desired degranulation outputs. To generate synthetic two-category qualitative data, we performed the same procedure, but recorded only whether the noise-corrupted primary degranulation response was greater or less than the noise-corrupted secondary degranulation response. To generate synthetic three-category qualitative data, we followed the same procedure, but recorded that the primary and secondary responses were approximately equal if the difference between the two responses was less than a designated threshold, which was set at $4.2\times 10^4$ arbitrary units. 

We performed MCMC sampling using PyBioNetFit's parallel tempering algorithm. For each dataset considered, we performed four independent runs and combined all samples obtained. Each run consisted of four Markov chains for each of nine temperatures, for a total of 36 chains, with samples saved from the four chains at temperature 1, run for a total of 50,000 steps including an unsampled 10,000-step burn-in period. Each run was performed using all 36 cores of a single Intel Broadwell E5-2695 v4 cluster node. Complete configuration settings are provided in the PyBioNetFit configuration file online.

\section{Results}

\subsection{Mathematical derivation}

\subsubsection{Notation}
\label{sec:problem}

By way of introduction to our newly proposed likelihood function for qualitative data, we begin by reviewing Bayesian UQ and its associated likelihood function with a more conventional quantitative dataset.

We are given an experimental dataset $\mathbf{y}=\{y_1,...,y_n\}$ and a model $f$. There is no restriction on what type of numerical measurement each $y_i$ represents; for example, it could represent a single data point of a time course, a sample mean of several independent and identically distributed measurements, or an arbitrary function of multiple measured quantities. Within a Bayesian framework, the $y_i$ are taken to be samples from the random variables  $\{Y_1,...,Y_n\}$. The model $f$ takes as input a parameter vector $\btheta$ to predict the expected value of each data point $Y_i$, that is, $f_i(\btheta) = E(Y_i)$. $\btheta$ is the realization of the random variable $\mathbf{\Theta}$. $f$ is assumed to be deterministic (e.g., an ODE model). Stochastic models would require additional treatment that is beyond the intended scope of this study. 

In Bayesian UQ, parameter uncertainty is quantified by the posterior probability distribution $P(\btheta|\mathbf{y})$, the probability of a particular parameter set given the data. Markov chain Monte Carlo (MCMC) algorithms can be used to sample the posterior distribution using the fact that, by Bayes' law, $P(\btheta|\mathbf{y}) \propto P(\mathbf{y}|\btheta)P(\btheta)$. The change in the value of $P(\mathbf{y}|\btheta)P(\btheta)$ is used to determine whether a proposed move by the MCMC algorithm is accepted. $P(\btheta)$ is a user-specified distribution representing prior knowledge about the parameters. Therefore, an important prerequisite for performing Bayesian UQ is an expression for the \textit{likelihood}, $P(\mathbf{y}|\btheta)$. 

\subsubsection{Chi squared likelihood function}

When performing conventional Bayesian UQ using only quantitative data, a common choice of likelihood function (e.g., see \cite{Kozer2013} and \cite{Harmon2017}) is the chi squared function.

\begin{equation}
    -\log P(\mathbf{y}|\btheta) \propto \chi^2(\btheta) = \sum_{i=1}^n \frac{(y_i-f_i(\btheta))^2}{2\sigma_i^2}
    \label{eq:chisq}
\end{equation}

Here $\sigma_i$ is the standard deviation of the measurement $y_i$. If $y_i$ represents the sample mean of several independent trials, it is common to estimate $\sigma_i$ as the standard error of the mean. 

This likelihood function has a strong theoretical motivation. The underlying assumption is that each $Y_i$ has an independent Gaussian distribution with mean $f_i(\btheta)$ and standard deviation  $\sigma_i$. Then the probability of a single data point $y_i$ given $\btheta$ is

\begin{equation}
    P(y_i|\btheta) = \frac{1}{\sqrt{2\pi}\sigma_i} \exp(\frac{-(y_i-f_i(\btheta))^2}{2\sigma_i^2})
\end{equation}

Given that the $Y_i$ are independent, the probability of the complete dataset $\mathbf{y}$ given $\btheta$ is given by the product

\begin{equation}
    P(\mathbf{y}|\btheta) = \prod_{i=1}^n \frac{1}{\sqrt{2\pi}\sigma_i} \exp(\frac{-(y_i-f_i(\btheta))^2}{2\sigma_i^2})
    \label{eq:product}
\end{equation}

When performing MCMC sampling, we typically only need a value \textit{proportional to} $P(\mathbf{y}|\btheta)$ to calculate the ratio $P(\mathbf{y}|\btheta_1) / P(\mathbf{y}|\btheta_2)$ for two parameter sets $\btheta_1$ and $\btheta_2$. This ratio is used to determine, for example, the probability of transitioning from $\btheta_1$ to $\btheta_2$ in the Metropolis-Hastings algorithm. We therefore can ignore proportionality constants in Equation \ref{eq:product} that are independent of $\btheta$. 

\begin{equation}
    P(\mathbf{y}|\btheta) \propto \prod_{i=1}^n \exp(\frac{-(y_i-f_i(\btheta))^2}{2\sigma_i^2})
    \label{eq:propproduct}
\end{equation}

Taking the negative logarithm of Equation \ref{eq:propproduct} results in the conventional chi squared function (Equation \ref{eq:chisq}). Therefore, under the assumptions stated in this section, the chi squared function represents the kernel of the negative log likelihood and can be rigorously used in Bayesian UQ algorithms. 

\subsubsection{Likelihood function for qualitative data}

We now consider the situation in which the experimental data are qualitative. By qualitative data, we specifically mean observations that can be expressed as inequality constraints to be enforced on outputs of a model.

Our problem statement is nearly identical to that presented in Section \ref{sec:problem}, except we are no longer given the dataset $\mathbf{y}$. Instead, for each $Y_i$,  we are given a constant $c_i$, and told whether $y_i < c_i$ or $y_i > c_i$ was observed. $y_i$ is the sample generated from $Y_i$ and is never observed. $y_i < c_i$ (or $y_i > c_i$) is the observation, which has two possible outcomes. We explicitly write down the procedure to generate these qualitative observations from the $Y_i$, which we refer to as our \textit{sampling model}:

\begin{algorithm}[H]
To generate observation $i$, sample $y_i$ from $Y_i$ and report whether $y_i < c_i$ or $y_i > c_i$.
\end{algorithm}
\vspace{-12pt}

Without loss of generality, we assume all given observations have the form $y_i < c_i$. If some quantity $A$ yielded an observation $a > k$, we could set $Y_i=-A$ and $c_i=-k$. This form also supports the case of an inequality $A<B$ between two measured quantities, as we could set $Y_i=A-B$ and $c_i=0$.

To perform Bayesian analysis, we require an expression for the probability of observing $y_i < c_i$ for all $i$ (rather than observing $y_i > c_i$ for some $i$), given a parameter set $\btheta$. As shorthand, we will write this as $P(\mathbf{y}<\mathbf{c}|\btheta)$, where $\mathbf{y}$ is a vector of the $y_i$ and $\mathbf{c}$ is a vector of the $c_i$. 

Following the example of the chi squared likelihood function, we assume each $Y_i$ has a Gaussian distribution with a known standard deviation $\sigma_i$. The mean of the distribution is, as before, taken to be given by the model prediction $f_i(\btheta)$. With this distribution, the probability of observing $y_i < c_i$ is, by definition, given by the Gaussian cumulative distribution function (CDF). We will write the CDF of a Gaussian distribution with mean $\mu$ and standard deviation $\sigma$ evaluated at a point $x$ as $\textrm{cdf}(\mu,\sigma,x)$. The conditional probability of interest is as follows:

\begin{equation}
    P(y_i<c_i|\btheta) = \textrm{cdf}(f_i(\btheta),\sigma_i,c_i) 
    \label{eq:qualsingle}
\end{equation}

We note that for ease of implementation, $\textrm{cdf}(\mu,\sigma,x)$ can be written in terms of the error function $\textrm{erf}(x)$, which is implemented in many standard libraries, including the Python and C++ standard libraries. 

\begin{equation}
    \textrm{cdf}(\mu,\sigma,x) = \mu + \frac{1 + \textrm{erf}(\frac{x}{\sigma \sqrt{2}})}{2}
\end{equation}

\begin{figure}[tb]
\centering\includegraphics[scale=0.5]{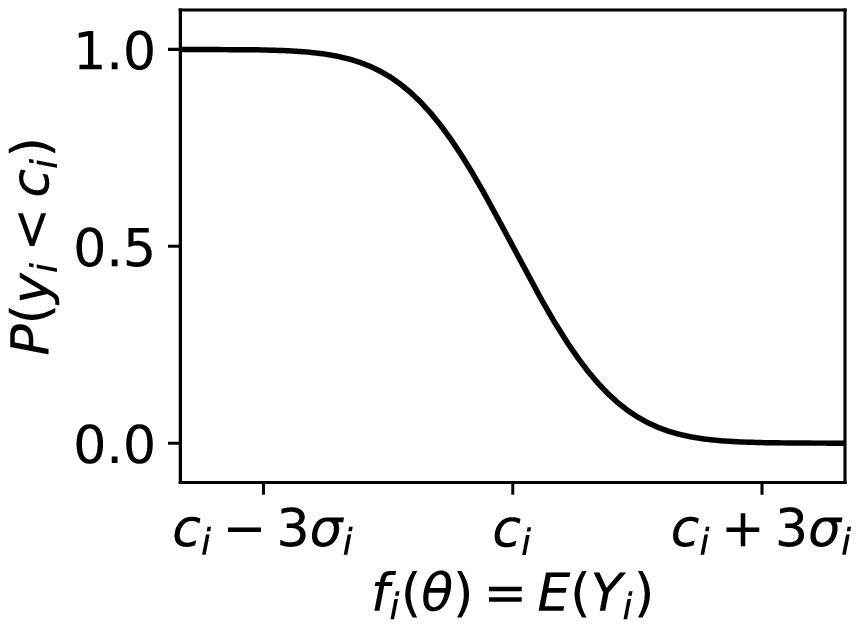}
\caption{The proposed form for $P(y_i<c_i|\btheta)$ (Equation \ref{eq:qualsingle}). }
\label{fig:logistic}
\end{figure}

As shown in Figure \ref{fig:logistic}, Equation \ref{eq:qualsingle} is intuitively reasonable. If the true mean value of $Y_i$ is much smaller than $c_i$ (relative to the scale of $\sigma_i$), we are very likely to observe $y_i<c_i$, whereas if the mean of $Y_i$ is much larger than $c_i$, we are very unlikely to observe $y_i<c_i$. If the true mean of $Y_i$ is close to $c_i$, we are uncertain whether the observation will be $y_i<c_i$ or $y_i>c_i$ in the face of measurement noise. We note that this function has a similar appearance to the logistic function, which is commonly used to model binary categorization in machine learning.

Assuming independence of the $Y_i$, the probability of the entire dataset is given by the product.

\begin{equation}
    P(\mathbf{y}<\mathbf{c}|\btheta) = \prod_{i=1}^n \textrm{cdf}(f_i(\btheta),\sigma_i,c_i)
    \label{eq:qualproduct}
\end{equation}

Finally, we take the negative logarithm to obtain

\begin{equation}
    -\log P(\mathbf{y}<\mathbf{c}|\btheta) = \sum_{i=1}^n -\log \textrm{cdf}(f_i(\btheta),\sigma_i,c_i) 
    \label{eq:qualobj}
\end{equation}

This function can be used for Bayesian UQ when considering qualitative data in an equivalent way to how the chi squared likelihood function is used when considering quantitative data. 




\subsubsection{Likelihood function for qualitative data with model discrepancy}

\label{sec:twocat}

The likelihood function in Equation \ref{eq:qualobj} has a remaining limitation when it comes to real-world experimental data. To illustrate this concern, we point to the model developed by Tyson and co-workers of yeast cell cycle control \citep{Chen2000,Chen2004,Csikasz-Nagy2006,Oguz2013,Kraikivski2015}. Several versions of this model have been parameterized using qualitative data (viability status of yeast mutants) by hand-tuning \citep{Chen2000,Chen2004,Csikasz-Nagy2006,Kraikivski2015} and with optimization algorithms \citep{Oguz2013,Mitra2018a}. In all of these parameterization studies, most but not all of the qualitative observations were satisfied by the reported best-fit parameterization. A few of the observations, however, were different from the model predictions. Due to such anomalous observations, a likelihood model as we have described could give the dataset a very low likelihood given the model and parameters, even though there intuitively is good agreement between the parameterized model and dataset.  

How can we reconcile anomalous observations? An explanation given by Tyson and co-workers is that a model has a limited amount of detail, which is unable to capture every qualitative observation in the data \citep{Chen2004}. This explanation suggests using a statistical approach known as model discrepancy or model inadequacy \citep{Kennedy2001}. The principle of model discrepancy is that when calculating the likelihood of a dataset, one should take into account the difference between the model and reality. Although many statistical studies ignore model discrepancy, it has been shown to be important for performing effective statistical inference for certain problems \citep{Brynjarsdottir2014}. Given that qualitative data may be generated by high-throughput screening that could easily step outside the scope of a particular model, we believe model discrepancy is an especially important consideration for our applications. 

Existing treatments of model discrepancy often describe discrepancy with its own probability distribution, such as a Gaussian distribution that is autocorrelated in time \citep{Brynjarsdottir2014}. Such an approach, which uses an assumption that model discrepancy is correlated for similar observations, is hard to apply to our problem formulation in which the $Y_i$ are taken to be independent (possibly coming from different model outputs). Thus, we take a more generic approach of expressing model discrepancy as a constant probability $\epsilon_i$ for each qualitative observation. $\epsilon_i$ relates to the probability that a given observation is outside the scope of the model. We say that when an observation is made, there is a probability $\epsilon_i$ that $y_i < c_i$ is reported regardless of the expected value of $Y_i$ given by the model. Likewise, there is also a probability $\epsilon_i$ that $y_i > c_i$ reported regardless of $Y_i$. These statements can be formalized as part of our sampling model:

\begin{algorithm}[H]
To generate observation $i$, make a weighted random choice of one of the following possibilities:
\begin{itemize}
    \item With probability $1-2\epsilon_i$, sample $y_i$ from $Y_i$ and report whether $y_i<c_i$ or $y_i>c_i$
    \item With probability $\epsilon_i$, report $y_i<c_i$
    \item With probability $\epsilon_i$, report $y_i>c_i$
\end{itemize}
\end{algorithm}
\vspace{-12pt}


With this modification, we have the probability distribution 

\begin{equation}
    P(y_i<c_i|\btheta) = \epsilon_i + (1-2\epsilon_i) \textrm{cdf}(f_i(\btheta),\sigma_i,c_i)
\end{equation}

and the likelihood function

\begin{equation}
    -\log P(\mathbf{y}<\mathbf{c}|\btheta) = \sum_{i=1}^n -\log (\epsilon_i + (1-2\epsilon_i) \textrm{cdf}(f_i(\btheta),\sigma_i,c_i)) 
    \label{eq:qualobjfinal}
\end{equation}

Equation \ref{eq:qualobjfinal} gives our recommended form for a likelihood function incorporating qualitative data with two possible categorical outcomes ($y_i<c_i$ or $y_i > c_i$). We will refer to this function as the \textit{two-category likelihood function}. 

Note that although we introduced $\epsilon_i$ for dealing with model structure problems, it could also represent a shortcoming of our postulated Gaussian error model. For example, if an experimental instrument had some probability of reporting a false positive or negative, regardless of whether the mean of $Y_i$ is close to the threshold $c_i$, this non-Gaussian error could be accounted for by increasing the value of $\epsilon_i$.

\subsubsection{Likelihood function for ordinal data with more than two categories}
\label{sec:threecat}

We next derive a likelihood function for ordinal categorical data with more than two categories. For simplicity, we suppose an observation has three possible outcomes: $y_i<c_{i,1}$, $c_{i,1}<y_i<c_{i,2}$, and $y_i>c_{i,2}$, for constants $c_{i,1}$ and $c_{i,2}$. An example would be if we were making an ordinary qualitative observation ($y_i<c_i$ or $y_i>c_i$), but another possible outcome of the experiment is $y_i=c_i$ to within the experimental error. Then the cutoffs $c_{i,1}$ and $c_{i,2}$ could be chosen on either side of $c_i$ such that the outcome $c_{i,1}<y_i<c_{i,2}$ corresponds to $y_i$ within measurement error.

From the definition of the Gaussian CDF we have

\begin{equation}
    \label{eq:yltc1}
    P(y_i<c_{i,1}|\btheta) = 1-\textrm{cdf}(f_i(\btheta),\sigma, c_{i,1})
\end{equation}
\begin{equation}
    P(c_{i,1}<y_i<c_{i,2}|\btheta) = \textrm{cdf}(f_i(\btheta),\sigma_i, c_{i,1}) - \textrm{cdf}(f_i(\btheta),\sigma_i, c_{i,2})
    \label{eq:pmiddle}
\end{equation}
\begin{equation}
    \label{eq:ygtc2}
    P(y_i>c_{i,2}|\btheta) = \textrm{cdf}(f_i(\btheta),\sigma_i, c_{i,2})
\end{equation}

A simplification is possible under the assumption that $c_{i,1}$ and $c_{i,2}$ are far enough separated that for any $E(Y_i)$, at most two of the three categories have non-negligible probability. That is, if $E(Y_i)$ is close enough to $c_{i,2}$ that observing $y_i>c_{i,2}$ is a probable outcome, $E(Y_i)$ is also high enough above $c_{i,1}$ that observing $y_i<c_{i,1}$ has a probability close to zero. Thus, we assume that for all $\btheta$, either $\textrm{cdf}(f_i(\btheta),\sigma_i, c_{i,1}) = 1$ or $\textrm{cdf}(f_i(\btheta),\sigma_i, c_{i,2}) = 0$. This assumption is reasonable because if it were false, it would mean the experiment cannot reliably distinguish between the three categories, and so the data would be better analyzed as two-category data. With this assumption, Equation \ref{eq:pmiddle} can be rewritten as

\begin{equation}
    P(c_{i,1}<y_i<c_{i,2}|\btheta) = \textrm{cdf}(f_i(\btheta),\sigma_i, c_{i,1}) * (1 - \textrm{cdf}(f_i(\btheta),\sigma_i, c_{i,2}))
    \label{eq:pmiddle2}
\end{equation}

Note that Equation \ref{eq:pmiddle2} is equivalent to Equation \ref{eq:qualproduct} for two independent constraints $c_{i,1}<y_i$ and $y_i<c_{i,2}$ arising from two-category observations. This makes for a convenient implementation: rather than explicitly considering the two-sided observation $c_{i,1}<y_i<c_{i,2}$, we can rewrite the observation as two independent one-sided observations $c_{i,1}<y_i$ and  $y_i<c_{i,2}$ described by Equation \ref{eq:qualproduct}.

A modification to the two-category case is necessary when model discrepancy is included as in Equation \ref{eq:qualobjfinal}. Here, care must be taken to ensure that in the sampling model the probability of all possible outcomes sums to 1. For example, a reasonable sampling model for a three-category observation would be the following:

\begin{algorithm}[H]
To generate observation $i$, make a weighted random choice of one of the following possibilities:
\begin{itemize}
    \item With probability $1-3\epsilon_i$, sample $y_i$ from $Y_i$ and report whether $y_i<c_{i,1}$ or $c_{i,1}<y_i<c_{i,2}$ or $c_{i,2}<y_i$
    \item With probability $\epsilon_i$, report $y_i<c_{i,1}$
    \item With probability $\epsilon_i$, report $c_{i,1}<y_i<c_{i,2}$
    \item With probability $\epsilon_i$, report $c_{i,2}<y_i$
\end{itemize}
\end{algorithm}
\vspace{-12pt}

Recall that in Equation \ref{eq:qualobjfinal}, in the case of model discrepancy, the observation is equally likely to be $y_i>c_i$ or $y_i<c_i$ (each of these events is assumed to have probability $\epsilon_i$).  In contrast, using the above sampling model, it is half as likely to report $y_i<c_{i,1}$ (probability $\epsilon_i$) as to report $y_i>c_{i,1}$ (probability $2\epsilon_i$). 

We generalize Equation \ref{eq:qualobjfinal} to account for the case of three-category observations by allowing for two separate parameters. We define the positive discrepancy rate $\epsilon_i^+$ as the probability that a constraint in the data is satisfied regardless of $Y_i$, and the negative discrepancy rate $\epsilon_i^-$ as the probability a constraint is violated regardless of $Y_i$. For example, with the above sampling model, for the observation $c_{i,2}<y_i$, we would use $\epsilon_i^+ = \epsilon_i$ and $\epsilon_i^- = 2\epsilon_i$

Our modified likelihood function is

\begin{equation}
    -\log P(\mathbf{y}<\mathbf{c}|\btheta) = \sum_{i=1}^n -\log (\epsilon_i^+ + (1-\epsilon_i^+-\epsilon_i^-) \textrm{cdf}(f_i(\btheta),\sigma_i,c_i)) 
    \label{eq:qualobjfinalplus}
\end{equation}

We will refer to this function as the \textit{many-category likelihood function}. 

The same formulation can be extended to allow for an arbitrary number of ordinal categories. For example, with four categories defined by the thresholds $c_{i,1}$, $c_{i,2}$, and $c_{i,3}$, we could write expressions analogous to Equations \ref{eq:yltc1}-\ref{eq:ygtc2} for $P(y_i<c_{i,1}|\btheta)$, $P(c_{i,1}<y_i<c_{i,2}|\btheta)$, $P(c_{i,2}<y_i<c_{i,3}|\btheta)$, and $P(y_i>c_{i,3}|\btheta)$.

\begin{figure}[tb!]
\centering\includegraphics{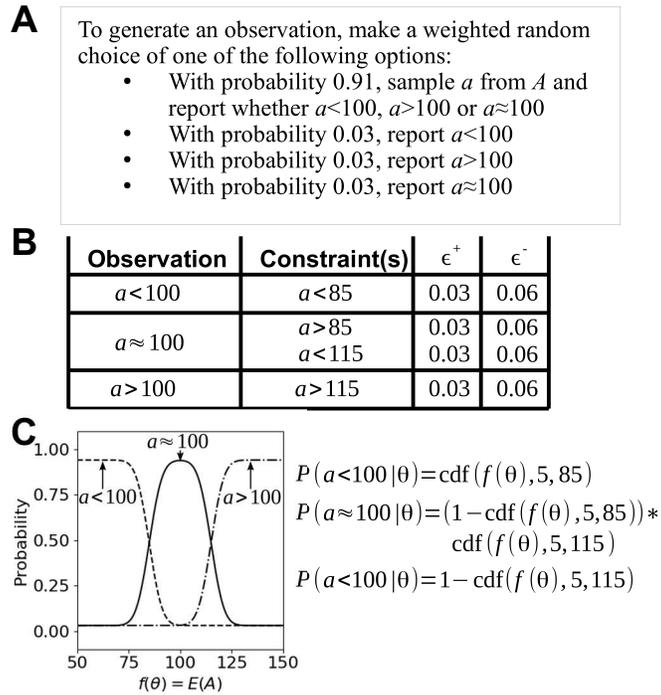}
\caption{Example constraints and probabilities arising from a qualitative observation with three possible categorical outcomes. (A) The sampling model associated with the observation. (B) Inequalities and $\epsilon^{+}$ and $\epsilon^{-}$ values associated with each possible observation outcome (C) Plots and equations giving the probability of each possible observation outcome as a function of the expected value of model output $A$. }
\label{fig:ex3}
\end{figure}

We illustrate the use of Equation \ref{eq:qualobjfinalplus} with a concrete example. Suppose we have a quantity of interest with the corresponding random variable $A$, and we make a qualitative observation with three possible outcomes: $a<100$, $a\approx100$, or $a>100$. Suppose also that based on the sensitivity of the assay, we know that any value of $a$ in the range 85--115 would be reported as ``$a\approx100$.'' Given this knowledge of the assay sensitivity, we take the standard deviation of $A$ to be 5, that is, we can only confidently report $a<100$ if $a$ is 3 standard deviations below the threshold of 100. We choose the sampling model shown in Figure \ref{fig:ex3}A, giving a base probability of 0.03 to each possible outcome due to model discrepancy. Note that this sampling model follows the requirement that the probabilities of all possible outcomes sum to 1. We then formulate the constraint(s) as shown in Figure \ref{fig:ex3}B, depending on whether the actual observation is $a<100$, $a\approx100$, or $a>100$. The resulting probabilities are shown in Figure \ref{fig:ex3}C as a function of the expected value of $A$ predicted by the model.

When using the many-category likelihood function, it is important to consider the underlying sampling model, and choose $\epsilon_i^+$ and $\epsilon_i^-$ in a way such that the probabilities in the sampling model sum to 1. An example of how to correctly choose $\epsilon_i^+$ and $\epsilon_i^-$ given a sampling model is presented in Section \ref{sec:application}.

\subsubsection{Combined likelihood function}

If independent quantitative and qualitative data are available, it is straightforward to combine the chi squared likelihood function for quantitative data with one of the newly presented likelihood functions for qualitative data. One would simply sum Equations \ref{eq:chisq} and \ref{eq:qualobjfinal} (or \ref{eq:qualobjfinalplus}) to obtain the kernel of the negative log likelihood for the combined dataset.

The relative weighting of the two datasets is determined by the standard deviations for the quantitative data points and the values of $\sigma_i$ and $\epsilon_i$ for the qualitative observations. 

\subsection{Software implementation}

We implemented the likelihood functions described in the previous section in PyBioNetFit v1.1.0. PyBioNetFit supports both the two-category (Equation \ref{eq:qualobjfinal}) and many-category (Equation \ref{eq:qualobjfinalplus}) likelihood functions for qualitative data, and supports combining these functions with the chi squared likelihood function for quantitative data. 

The new options were added via an extension of the Biological Property Specification Language (BPSL) supported by PyBioNetFit. As previously described \citep{Mitra2019a}, a BPSL statement consists of an inequality, followed by an enforcement condition, followed by a weight. For example, in the statement
\begin{equation*}
    \texttt{A<4 at time=1 weight 2}
\end{equation*}
the inequality is \texttt{A<4} (referring to some modeled quantity $A$), the enforcement condition is \texttt{time=1} (referring to time 1 in a time course), and the weight is declared by \texttt{weight 2}. This weight declaration refers to $C_i$ in the previously described static penalty function (Equation \ref{eq:static}). Using this formulation, the term added to an objective function for this constraint would be $2 \cdot \max(0,A(1)-4)$, where $A(1)$ is model output $A$ evaluated at time = 1. 

In PyBioNetFit v1.1.0, we added an alternative to the weight clause to specify parameters of the new likelihood functions. As described in Section \ref{sec:twocat}, for each inequality in the data, the two-category likelihood function has two user-configurable parameters: the probability $\epsilon_i$ of measuring $y_i<c_i$ regardless of the distribution of $Y_i$, and the standard deviation $\sigma_i$ of the quantity $Y_i$. The value of $1-2\epsilon_i$ (i.e., the probability that the distribution of $Y_i$ is relevant to the experimental result) is supplied to PyBioNetFit with the \texttt{confidence} keyword. $\sigma_i$ is supplied to PyBioNetFit with the \texttt{tolerance} keyword. Therefore, an example BPSL statement using the two-category likelihood function is

\begin{equation*}
    \texttt{A<4 at time=1 confidence 0.98 tolerance 0.5}
\end{equation*}

This statement would result in using the likelihood function of Equation \ref{eq:qualobjfinal} with $\epsilon_i=0.01$, $\sigma_i=0.5$, $c_i=4$, and $Y_i=A(1)$. The resulting term added to the likelihood function is $-\textrm{log}(0.01+0.98\cdot\textrm{cdf}(A(1),0.5,4))$.

PyBioNetFit also supports the use of the many-category likelihood function (Equation \ref{eq:qualobjfinalplus}) through the specification of separate positive and negative discrepancy rates. In this case, the \texttt{confidence} keyword is replaced with the keywords \texttt{pmin} to specify $\epsilon_i^-$ (i.e., the minimum value of $P(y_i<c_i|\btheta)$) and \texttt{pmax} to specify $1-\epsilon_i^+$ (i.e., the maximum value of $P(y_i<c_i|\btheta)$). For example, the BPSL statement

\begin{equation*}
    \texttt{A<4 at time=1 pmin 0.01 pmax 0.98 tolerance 0.5}
\end{equation*}
would use Equation \ref{eq:qualobjfinalplus} with $\epsilon_i^+=0.02$, $\epsilon_i^-=0.01$, $\sigma_i=0.5$, $c_i=4$, and $Y_i=A(1)$. The resulting term added to the likelihood function is $-\textrm{log}(0.01+0.97\cdot\textrm{cdf}(A(1),0.5,4))$.

When writing these statements in BPSL, care must be taken to ensure that results are statistically valid. First, note that the \texttt{tolerance} specifies the standard deviation of the final random variable $Y_i$ used to sample $y_i$ in Equation \ref{eq:qualobjfinal}. For example in the above statement, it refers to the standard deviation of $A(1)$. In the statement \texttt{A>B at time=5 confidence 0.98 tolerance 0.5}, \texttt{tolerance} refers to the standard deviation of $A(5)-B(5)$, i.e., the sum of the standard deviations of $A(5)$ and $B(5)$. In the statement \texttt{A>4 always confidence 0.98 tolerance 0.5}, \texttt{tolerance} refers to the standard deviation of $\min(A(t))$, rather than the value of $A$ at any particular time.

Second, it is important to keep in mind the underlying sampling model to correctly set \texttt{confidence} or \texttt{pmin} and \texttt{pmax}. For example, in the sampling model of Fig \ref{fig:ex3}A, there are three possible constraints each with probability 0.03 to be satisfied due to model discrepancy and probability 0.06 to be violated due to model discrepancy. Therefore, the correct setting is \texttt{pmin 0.03 pmax 0.94}. 

Third, when using PyBioNetFit's enforcement keywords \texttt{always}, \texttt{once}, and \texttt{between}, it is important to be sure the possible categories in the sampling model are mutually exclusive and cover all possible outcomes. For example, if one of two possible categorical outcomes is \texttt{A>4 always}, the other must be \texttt{A<4 once} (not \texttt{A<4 always}). Likewise, if one category is \texttt{A>4 between time=5,time=10}, its negation is \texttt{A<4 once between time=5,time=10}. We note that the \texttt{once between} enforcement condition used here is a new feature of BPSL introduced in PyBioNetFit v1.1.1.

The sampling model is never explicitly input into PyBioNetFit, as equations \ref{eq:qualobjfinal} and \ref{eq:qualobjfinalplus} are defined regardless of whether the sampling model is well-defined. It is the user's responsibility to choose a well-defined sampling model and specify constraints accordingly to obtain meaningful results. 

\begin{figure}[tbp!]
\centering\includegraphics{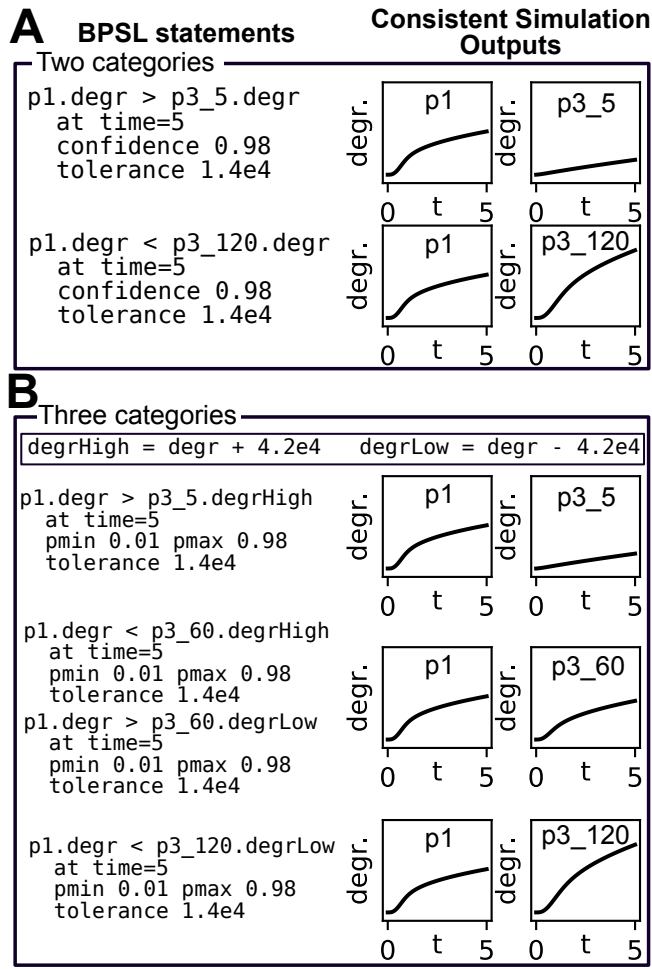}
\caption{Configuration of the example problem in BPSL. As described in the text, we considered the problem assuming either (A) two possible observation categories or (B) three possible categories. The left column shows an example BPSL statement for each possible category. In these BPSL statements, \texttt{p1} refers to the primary degranulation and \texttt{p3\_}$<t>$ refers to the secondary degranulation after a delay of $t$ minutes. Note that in the three-category case, the middle category requires two separate BPSL statements. The right column shows simulated trajectories of the primary (left) and secondary (right) degranulation responses that are consistent with the BPSL statement. For the three-category case, \texttt{degrHigh} and \texttt{degrLow} are functions defined in the BNGL model file for use in the BPSL statements.}
\label{fig:setup}
\end{figure}

\subsection{Example application}
\label{sec:application}

To demonstrate the use of qualitative likelihood functions in PyBioNetFit, we performed Bayesian UQ on a synthetic example problem based on the study of \cite{Harmon2017}. The model of \cite{Harmon2017} describes the degranulation of mast cells in response to two consecutive stimuli with multivalent antigen. In the original study, it was found that depending on the time delay between the two stimuli, the secondary response could be either stronger or weaker than the primary response. The original data consisted of quantitative degranulation measurements for six different time delays.

In our synthetic problem, we suppose that the experimental data took a different form. Rather than quantitative measurements, we assume that it is only possible to measure whether the secondary degranulation is higher or lower than the primary degranulation. These measurements can be seen as case-control comparisons between several conditions of interest (secondary degranulation at various time delays) and a control (primary degranulation). We assume that these measurements can be made at a larger number of time delays than were used in the original study (i.e., we have a less precise but higher throughput instrument than in the actual study). 

We generated synthetic data of this form using the published parameter values of the model as ground truth. For each time delay in the data, we ran a simulation, and added Gaussian noise to the primary and secondary degranulation outputs before recording whether the primary or secondary was higher. We generated datasets ranging from 4 to 64 time delays. The resulting datasets were implemented in BPSL as illustrated in Figure \ref{fig:setup}A. Note that we set the \texttt{confidence} to 0.98, allowing for a 0.02 chance of model discrepancy (although there is no true model discrepancy in this synthetic problem). We set the \texttt{tolerance} to $1.4 \times 10^4$, which is the standard deviation of the difference between the primary and secondary degranulation values (i.e., twice the standard deviation of the added noise for each individual degranulation value).

\begin{figure*}[!!!!!tbh!!!!!!!!!!!!]
\centering\includegraphics{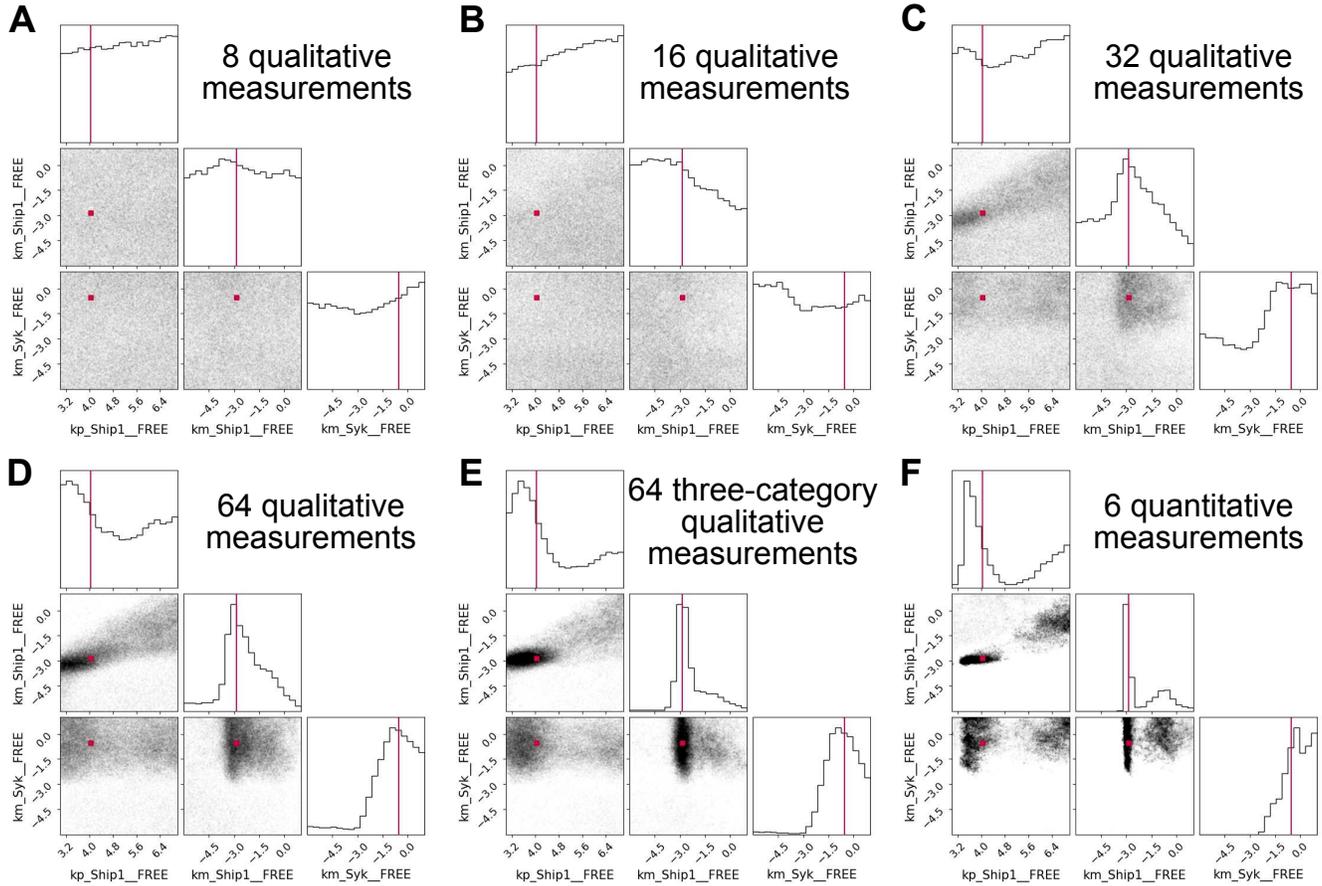}
\caption{Posterior distributions calculated by parallel tempering for three selected model parameters under different measurement protocols. (A-D) Datasets consisted of (A) 8, (B) 16, (C) 32, (D) 64 qualitative observations, each with two possible categorical outcomes. (E) The dataset consisted of 64 qualitative observations, each with three possible categorical outcomes. Results for datasets of 4, 8, 16, and 32 measurements are provided in Supplementary information. (F) The dataset consisted of six quantitative data points, similar to in the original study of \cite{Harmon2017}. Ground truth values are marked in red. The posterior distributions of all parameters are provided in Supplementary information. }
\label{fig:bayes}
\end{figure*}

We configured PyBioNetFit jobs to perform Bayesian UQ by parallel tempering for each dataset. The results are shown in Figure \ref{fig:bayes}A-D and Figures S1--S5. Not surprisingly, as the number of qualitative observations increases, we obtain a narrower distribution of parameter values, and these narrower distributions include the ground truth parameter values. This result demonstrates that with a sufficient amount of qualitative data, it is possible to find nontrivial credible intervals for parameter values.

To demonstrate the use of the many-category likelihood function (Equation \ref{eq:qualobjfinalplus}), we repeated the analysis using three-category synthetic data. Our three categories allow the secondary degranulation to be measured as smaller, larger, or within error of the primary degranulation. The three-category dataset was declared in BPSL as illustrated in Figure \ref{fig:setup}B. Compared to the two-category synthetic data, modifications were required as described in Section \ref{sec:threecat}. The assumed sampling model used for the constraints in Figure \ref{fig:setup}B is the following, where $Y_i$ represents the primary degranulation minus the secondary degranulation:

\begin{algorithm}[H]
To generate observation $i$, make a weighted random choice of one of the following possibilities:
\begin{itemize}
    \item With probability 0.97, sample $y_i$ from $Y_i$ and report whether $y_i<-4.2\times 10^4$ or $-4.2\times 10^4<y_i<4.2\times 10^4$ or $4.2\times 10^4<y_i$
    \item With probability 0.01, report $y_i<-4.2\times 10^4$
    \item With probability 0.01, report $-4.2\times 10^4<y_i<4.2\times 10^4$
    \item With probability 0.01, report $4.2\times 10^4<y_i$
\end{itemize}
\end{algorithm}
\vspace{-12pt}

We have chosen a threshold of $4.2\times 10^4$ for the difference between primary and secondary degranulation that qualifies as ``within error.'' This value is three times the standard deviation of $Y_i$, giving the separation of categories required in Section \ref{sec:threecat} (i.e., any sampled $y_i$ is consistent with at most two possible categories). This condition allows us to define the middle category ($-4.2\times 10^4<y_i<4.2\times 10^4$) using two independent BPSL statements. The choice of threshold is reflected in the BPSL by the use of the model outputs referred to as \texttt{degrHigh} and \texttt{degrLow}. Based on the sampling model, each category has a minimum probability of 0.01 due to model discrepancy, and a maximum probability of 0.98 (because the other two categories each have a minimum of 0.01). Therefore, we set \texttt{pmin} to 0.01 and \texttt{pmax} to 0.98 instead of using the \texttt{confidence} keyword. Finally, the \texttt{tolerance} is set to $1.4\times 10^4$, the same as for the two-category dataset. 

The results of parallel tempering using this dataset are illustrated in Figure \ref{fig:bayes}E and Figures S6--S10. As expected, compared to the results with two-category dataset of the same size, some parameters are bounded more tightly around their ground truth values. 

For comparison, we also performed the analysis using synthetic quantitative data generated at the same time delays as in the original study (Figure \ref{fig:bayes}F and Figure S11). The quantitative dataset produced distributions even tighter than those of the three-category qualitative data. It is notable how close we can get to the results with quantitative data by using purely qualitative data. 

\section{Discussion}

Here we have presented a new statistical framework for using qualitative data in conjunction with Bayesian UQ for biological models. In these models, unidentifiable parameters are common, but Bayesian analysis can determine which parameters and correlations are identifiable, and to what extent the model has predictive value despite unidentifiable parameters. 

We see this framework as a more statistically rigorous improvement upon our previously described static penalty function approach \citep{Mitra2018a} (Equation \ref{eq:static}). Our new framework can be used for statistical analysis, whereas the previous formulation was simply a heuristic for finding a single reasonable parameter set.  

Our new likelihood function has applications beyond Bayesian UQ. It can also, like the static penalty function, be used with optimization algorithms to find a point estimate of the best parameters. In such a problem, the global minimum (assuming it can be found by an optimization algorithm) is the maximum likelihood estimate, i.e., the maximum of the posterior distribution. The new likelihood function may also be used for UQ by profile likelihood analysis \citep{Kreutz2013}.

The static penalty function may remain more efficient at point estimation. The cdf-based likelihood function has the limitation that when far from constraint satisfaction, its gradient is near zero, and so it cannot effectively guide the optimization algorithm toward constraint satisfaction. In contrast, the static penalty function provides useful information for optimization at any distance from constraint satisfaction. One potential workflow could be to use the static penalty function for initial optimization, followed by the likelihood function for refinement and evaluation of the best fit. 

We note that under our new framework, each constraint now has two adjustable settings: $\epsilon_i$ and $\sigma_i$. This may appear worse than the single weight parameter $C_i$ in the static penalty formulation, but the advantage is that both of these parameters have a statistical interpretation. $\epsilon_i$ represents the probability of model discrepancy resulting in a qualitative observation that occurs regardless of the model and its predicted mean. $\sigma_i$ represents the standard deviation of the quantity considered in the constraint. This value might seem challenging to estimate, given we may not even be able to quantitatively measure the quantity of interest. 
However, much of the same intuition holds as when dealing with Gaussian-distributed quantitative data. In particular, if there is a difference of $2\sigma_i$ between a threshold and the mean, we can be reasonably confident (probability 97.7\%) that an observation would yield the correct result (greater or less than the threshold). With a difference of $3\sigma_i$, we can be extremely confident (probability 99.87\%). 
To choose $\sigma_i$, a reasonable thought process would be to ask, ``How large of a difference would there have to be for the experiment to be sure to detect the difference?'', and set $\sigma_i$ equal to one third of that difference.

Both parameters can be seen as optional. If we don't expect a scenario in which a constraint is impossible to reconcile with our model, we can set $\epsilon_i=0$, ignoring this aspect of the likelihood function. Likewise, if we have no way to estimate the standard deviation of the measured quantity, we could set $\sigma_i=0$ and use $\epsilon_i$ to set a fixed probability of satisfying the constraint. Thus, the two adjustable constants should be seen as an opportunity to provide all available information about a qualitative observation of interest, rather than as a burden for manual adjustment. 

We expect that our new formulation of a likelihood function derived from qualitative data will be useful in future modeling studies and will help facilitate the wider adoption of qualitative data as a data source for model parameterization.


\section*{Acknowledgements}

We acknowledge computational resources provided by the Institutional Computing program at Los Alamos National Laboratory, which is operated by Triad National Security, LLC for the NNSA of DOE under contract 9233218CNA000001. We thank Steven Sanche for useful discussions.\vspace*{-12pt}

\section*{Funding}

This work has been supported by NIH/NIGMS grant R01GM111510.\vspace*{-12pt}

\bibliography{references}

\end{document}